\documentclass[%
 reprint,
superscriptaddress,
amsmath,amssymb,
aps,
pra,
]{revtex4-1}
\usepackage{caption}
\usepackage{subcaption}
\usepackage{graphicx}
\usepackage{xcolor}
\usepackage{graphicx}
\usepackage{dcolumn}
\usepackage{bm}

\begin{document}

\title{Observation of ultrabroadband striped space-time surface plasmon polaritons}

\preprint{APS/123-QED}

\author{Naoki Ichiji}
\affiliation{Graduate School of Pure and Applied Sciences, University of Tsukuba, 1-1-1 Tennodai, Tsukuba-shi, Ibaraki 305-8571, Japan}
\author{Hibiki Kikuchi}
\affiliation{School of Science and Engineering, University of Tsukuba, 1-1-1 Tennodai, Tsukuba-shi, Ibaraki 305-8571, Japan}
\author{Murat Yessenov}
\affiliation{CREOL, The College of Optics \& Photonics, University of Central Florida, Orlando, FL 32816, USA}
\author{Kenneth L. Schepler}
\affiliation{CREOL, The College of Optics \& Photonics, University of Central Florida, Orlando, FL 32816, USA}
\author{Ayman F. Abouraddy}
\thanks{raddy@creol.ucf.edu}
\affiliation{CREOL, The College of Optics \& Photonics, University of Central Florida, Orlando, FL 32816, USA}
\author{Atsushi Kubo}
\email{kubo.atsushi.ka@u.tsukuba.ac.jp}
\affiliation{Faculty of Pure and Applied Sciences, University of Tsukuba,1-1-1 Tennodai, Tsukuba-shi, Ibaraki 305-8571, Japan}


\begin{abstract}
Because surface plasmon polaritons (SPPs) are surface waves characterized by one free transverse dimension, the only monochromatic diffraction-free spatial profiles for SPPs are cosine and Airy waves. Pulsed SPP wave packets have been recently formulated that are propagation-invariant and localized in the in-plane dimensions by virtue of a tight spectral association between their spatial and temporal frequencies, which have thus been dubbed `space-time' (ST) SPPs. Because of the spatio-temporal spectral structure unique to ST-SPPs, the optimal launching strategy of such novel plasmonic field configurations remains an open question. We present here a critical step towards realizing ST-SPPs by reporting observations of ultrabroadband striped ST-SPPs. These are SPPs in which each wavelength travels at a prescribed angle with respect to the propagation axis to produce a periodic (striped) transverse spatial profile that is diffraction-free. We start with a free-space ST wave packet that is coupled to a ST-SPP at a gold-dielectric interface, and unambiguously identify the ST-SPP via an axial beating detected in two-photon fluorescence produced by the superposition of incident ST wave packet and the excited surface-bound ST-SPP. These results highlight a viable approach for efficient and reliable coupling to ST-SPPs, and thus represent the first crucial step towards realization of the full potential of ST-SPPs for plasmonic sensing and imaging. 
\end{abstract}

\maketitle

\section{Introduction}

The strong localization of surface plasmon polaritons (SPPs) at metal-dielectric interfaces has enabled a broad swath of applications in sensing \cite{Anker08NM,Homola08CR}, superresolution imaging \cite{Fang05Science,Kawata08NP,Willets17CR,Lee20NL}, optical tweezers \cite{Righini08PRL,Roxworthy12NL,Zhang21Light}, and information processing \cite{MacDonald09NP,Melikyan11OE,Melikyan14NP,Ono20NP}. However, in absence of a transverse confining structure \cite{Bozhevolnyi06Nature,Oulton08NP,Fang15Light}, SPPs diffract in the transverse dimension just as free optical fields do in the bulk. Furthermore, because SPPs have only one free transverse dimension, the so-called `diffraction-free' beam structures that have proved useful in free space \cite{Levy16PO} (e.g., Bessel \cite{Durnin87PRL} and Matthieu \cite{Gutierrez03AJP} beams) are not a viable option. These monochromatic beams require \textit{two} transverse dimensions to resist diffraction; e.g., in contrast to its two-dimensional (2D) counterpart, the one-dimensional (1D) Bessel beam diffracts. Indeed, the only diffraction-free monochromatic optical fields in 1D are cosine and Airy waves \cite{Siviloglou07OL} -- both of which have been exploited to produce monochromatic SPPs (cosine-Gauss SPPs \cite{Lin12PRL} and Airy plasmons \cite{Salandrino10OL,Minovich11PRL,Zhang11OL,Li11PRL}).

Recently a new class of propagation-invariant \textit{pulsed} beams (diffraction-free \textit{and} dispersion-free) have been investigated under the general rubric of `space-time' (ST) wave packets, which require introducing a tight spectral association between the wavelengths and the spatial frequencies that conforms to a prescribed functional form \cite{Kondakci16OE,Parker16OE,Kondakci17NP,Porras17OL,Efremidis17OL,Wong17ACSP2,Yessenov19OPN,Yessenov22AOP}. Besides propagation invariance in free space \cite{Turunen10PO,FigueroaBook14,Bhaduri19OL}, as well as in non-dispersive \cite{Bhaduri19Optica} or dispersive \cite{Longhi04OL,Porras04PRE,Malaguti08OL,Malaguti09PRA,Hall22APLP,Bejot22arxiv} dielectrics, ST wave packets feature a host of unique characteristics \cite{Yessenov19OPN,Yessenov22AOP}, including tunable group velocities \cite{Salo01JOA,Kondakci19NC} and anomalous refraction \cite{Bhaduri20NP,Motz21OL}. A critical feature of ST wave packets makes them particularly useful candidates for SPPs: their diffraction-free behavior is independent of dimensionality \cite{Yessenov22AOP} (whether one \cite{Kondakci17NP} or two \cite{Pang2021OL,Yessenov22NC} transverse spatial coordinates are involved), making them excellent candidates for plasmonic applications that require maintaining transverse spatial localization in sensing and imaging.

\begin{figure}[t!]
  \begin{center}
  \includegraphics[width=8.6cm]{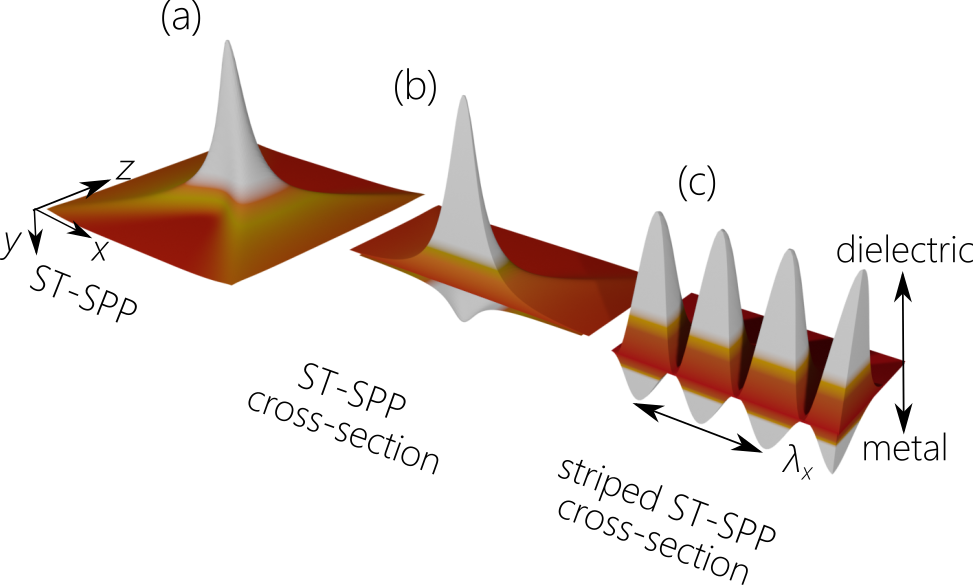}
  \end{center}
  \caption{Concept of striped ST-SPPs. (a) Illustration of a ST-SPP that is localized along $y$ by virtue of intrinsic plasmonic confinement, along $x$ because of the finite spatial bandwidth (or beam width), and along $z$ because of the finite temporal bandwidth (or pulse width). Propagation invariance is maintained by virtue of the spatio-temporal spectral structure inculcated into the SPP field [Fig.~\ref{Fig:LightConeRepresentation}(d)]. (b) A section through the ST-SPP at its mid-point along $z$ that highlights its transverse localization along $x$ and $y$ at a fixed axial plane $z$. The field extends into the dielectric further than in the metal. (c) Illustration of a cross-section through a striped ST-SPP with transverse period $\lambda_{x}$.}
  \label{Fig:GeneralConcept}
\end{figure}

We recently investigated the propagation characteristics of pulsed SPPs having a finite transverse extent in the generic setting of a metal-dielectric interface \cite{Schepler20ACSP}. By structuring the spatio-temporal spectra of these surface waves in a similar manner to ST wave packets, we verified theoretically that they are localized in all dimensions (the in-plane dimensions, in addition to the out-of-plane plasmonic confinement), are propagation invariant (the transverse profile resists diffraction, and the pulse profile resists dispersion independently of the materials selected), and their group velocity can be tuned from subluminal to superluminal values \cite{Schepler20ACSP}. We have thus dubbed these surface-bound wave packets ST-SPPs [Fig.~\ref{Fig:GeneralConcept}(a,b)]. However, ST-SPPs have \textit{not} been realized experimentally to date. Indeed, the question regarding the optimal strategy for launching ST-SPPs at a metal-dielectric interface from a free-space ST wave packet \textit{without} modifying its spatio-temporal spectral structure remains an open question. Whereas conventional approaches such as grating couplers succeeded in launching cosine and Airy plasmons in the monochromatic regime, they are not appropriate for ST-SPPs -- because they can alter the spectral structure of the incident field that is key to the unique propagation characteristics of ST-SPPs. Furthermore, care must be taken to clearly ascertain that a surface-bound wave has been excited with the targeted characteristics, and to distinguish it from the incident free-space ST wave packet.

Here we take the first crucial step towards realizing ST-SPPs by launching a space-time-coupled ultrabroadband pulsed laser beam into a SPP at a gold-dielectric interface via a nano-scale slit \cite{Kubo07NL,Zhang11PRB,Zhang13JPCC}. The pulsed laser has a FWHM-bandwidth of $\Delta\lambda\!=\!110$~nm (10-fs pulse duration) at a center wavelength of $\lambda_{\mathrm{o}}\!=\!800$~nm. By directing each wavelength in free space at a different angle with respect to a fixed propagation axis, we incorporate angular dispersion \cite{Torres10AOP} into the field to hold the spatial frequency fixed across the entire bandwidth. This free-space optical field is then coupled to a surface-bound wave via a 100-nm-wide nano-slit that is ion-milled into the gold (Au). This particular excitation mechanism preserves the transverse spatial frequency of the incident field in the coupled SPP. Because a fixed spatial frequency is maintained in the incident optical field, this approach produces a `striped' ST-SPP having a periodic transverse profile, whose period is tuned from 10~$\mu$m to 30~$\mu$m [Fig.~\ref{Fig:GeneralConcept}(c)]. The striped ST-SPP excited on the Au surface is microscopically detected by imaging the two-photon florescence emitted from a 25-nm-thick dye-doped polymer layer coating the Au surface, which reveals distinctive beat patterns along both the transverse and the axial directions -- each of which is modulated with an individual periodicity. Along the transverse direction, the fluorescence image shows a periodicity equal to a half of that of a strived ST-SPP, reflecting two antinodes incorporated in one cycle of the transverse profile. The axial beating is the result of interference between the free-space ST wave packet and the surface-bound striped ST-SPP \cite{Hattori12JJAP,Ichiji22NanoP}. These observations provide evidence for the feasibility of synthesizing arbitrary ST-SPPs that incorporate a multiplicity of spatial frequencies \cite{Schepler20ACSP}.

\section{Theoretical formulation}

\subsection{Theory of ST wave packets}

\begin{figure}[t!]
  \begin{center}
  \includegraphics[width=8.6cm]{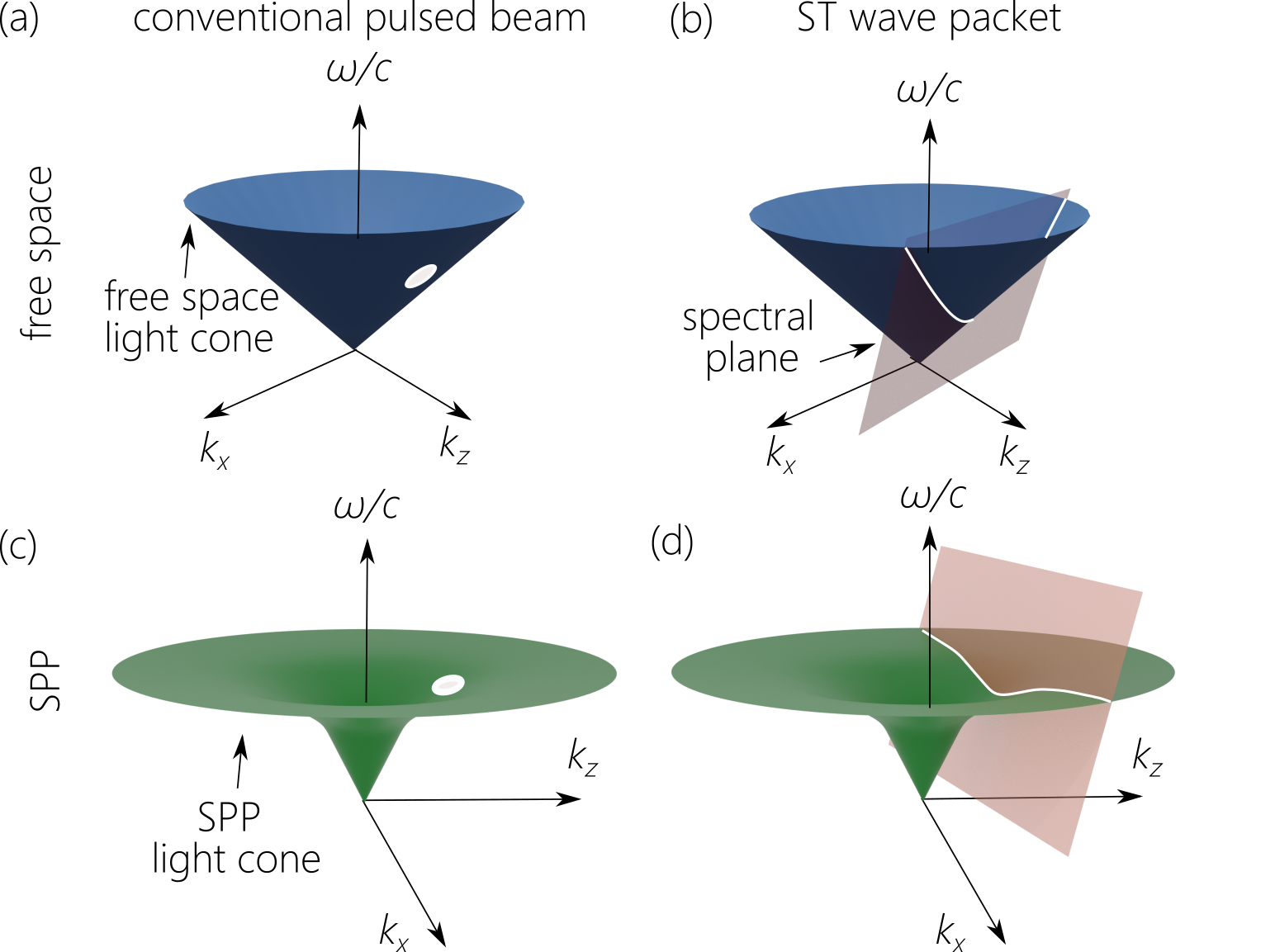}
  \end{center}
  \caption{Spectral representation of pulsed beams on the surface of the light-cone in $(k_{x},k_{z},\tfrac{\omega}{c})$-space. (a) Spectral support domain of a conventional pulsed beam, and (b) a ST wave packet on the surface of the free-space light-cone $k_{x}^{2}+k_{z}^{2}\!=\!(\tfrac{\omega}{c})^{2}$. (c) Spectral support domain on the surface of the SPP light-cone of a conventional pulsed SPP with finite spatial profile, and (d) a ST-SPP \cite{Schepler20ACSP}.}
  \label{Fig:LightConeRepresentation}
\end{figure}

For our purpose here, it is particularly useful to visualize pulsed optical beams in terms of the representation of their spectral support domain on the surface of the light-cone \cite{Donnelly93ProcRSLA,Yessenov19PRA,Yessenov22AOP}. In free space, if we retain one transverse coordinate $x$ (in anticipation of the formulation of ST-SPPs) along with the axial coordinate $z$, the dispersion relationship $k_{x}^{2}+k_{z}^{2}\!=\!(\tfrac{\omega}{c})^{2}$ defines the light-cone surface [Fig.~\ref{Fig:LightConeRepresentation}(a)]; here $k_{x}$ and $k_{z}$ are the transverse and axial wave numbers along $x$ and $z$, respectively, $\omega$ is the temporal frequency, and $c$ is the speed of light in vacuum. The spectral support domain for a conventional pulsed beam -- in which the spatial and temporal degrees of freedom are separable -- takes the form of a 2D region [Fig.~\ref{Fig:LightConeRepresentation}(a)], indicating finite spatial and temporal bandwidths \cite{Yessenov19PRA}.

In contrast, the spectral support domain for a ST wave packet in free space  -- while featuring finite spatial and temporal bandwidths -- is the 1D trajectory at the intersection of the light-cone with a tilted spectral plane \cite{Kondakci17NP}. This spectral trajectory determines the relationship between the spatial and temporal frequencies of the plane waves undergirding the field structure \cite{Donnelly93ProcRSLA,Yessenov22AOP}. The spectral plane is parallel to the $k_{x}$-axis [Fig.~\ref{Fig:LightConeRepresentation}(b)], and is defined as $k_{z}\!=\!k_{\mathrm{o}}+\Omega/\widetilde{v}$, where $\Omega\!=\!\omega-\omega_{\mathrm{o}}$, $\omega_{\mathrm{o}}$ is a fixed temporal frequency,  $k_{\mathrm{o}}\!=\!\tfrac{\omega_{\mathrm{o}}}{c}$ is the associated wave number, and $\widetilde{v}$ is the group velocity of the ST wave packet, which can in principle take on arbitrary values \cite{Salo01JOA,Wong17ACSP2,Kondakci19NC,Bhaduri19Optica}. Such a wave packet travels rigidly in free space at a group velocity $\widetilde{v}$ \cite{Turunen10PO,Kondakci17NP,Yessenov19OE}. Because each spatial frequency $k_{x}$ is associated with a single temporal frequency $\omega$, the spatial and temporal bandwidths, $\Delta k_{x}$ and $\Delta\omega$, respectively, are related \cite{Kondakci17NP,Yessenov19OE}: $\tfrac{\Delta\omega}{\omega_{\mathrm{o}}}\!=\!(\tfrac{\Delta k_{x}}{k_{\mathrm{o}}})^{2}\tfrac{1}{2|1-\widetilde{n}|}$, where $\widetilde{n}\!=\!c/\widetilde{v}$ is the group index of the free-space ST wave packet. We can then write $k_{x}(\omega)\!=\!\tfrac{\omega}{c}\sin{\{\varphi(\omega)\}}$ and $k_{z}(\omega)\!=\!\tfrac{\omega}{c}\cos{\{\varphi(\omega)\}}$, where $\varphi(\omega)$ is the propagation angle of the single frequency component $\omega$ with the $z$-axis, and $\sin\{\varphi(\omega)\}\!=\!\pm\sqrt{2(1-\widetilde{n})\tfrac{\Omega}{\omega_{\mathrm{o}}}+(1-\widetilde{n}^{2})(\tfrac{\Omega}{\omega_{\mathrm{o}}})^{2}}$; where $\Omega\!>\!0$ when $\widetilde{n}\!<\!1$ ($\omega\!>\omega_{\mathrm{o}}$ for superluminal ST wave packets), and $\Omega\!<\!0$ when $\widetilde{n}\!>\!1$ ($\omega\!<\!\omega_{\mathrm{o}}$ for subluminal ST wave packets) \cite{Yessenov19PRA}. Note that $\varphi(\omega)$ is \textit{non-differentiable} in the vicinity of $\omega\!=\!\omega_{\mathrm{o}}$ ($\Omega\!=\!0$), which is crucial for tuning the group velocity and maintaining propagation invariance \cite{Hall21OL,Hall21OL3NormalGVD,Hall22OEConsequences}. 
\subsection{Theory of ST-SPPs}

In contrast to the free-space light-cone $k_{x}^{2}+k_{z}^{2}\!=\!(\tfrac{\omega}{c})^{2}$, the light-cone for SPPs at the interface between a metal and a dielectric having relative permittivities $\epsilon_{\mathrm{m}}(\omega)$ and $\epsilon_{\mathrm{d}}(\omega)$, respectively, is given by $k_{x}^{2}+k_{z}^{2}\!=\!(\tfrac{\omega}{c})^{2}\tfrac{\epsilon_{\mathrm{m}}(\omega)\epsilon_{\mathrm{d}}(\omega)}{\epsilon_{\mathrm{m}}(\omega)+\epsilon_{\mathrm{d}}(\omega)}$. This SPP dispersion relation is enforced by localization along $y$ with wave number $k_{y,\mathrm{d}}\!=\!\tfrac{\omega}{c}\tfrac{\epsilon_{\mathrm{d}}(\omega)}{\sqrt{\epsilon_{\mathrm{d}}(\omega)+\epsilon_{\mathrm{m}}(\omega)}}$ in the dielectric and $k_{y,\mathrm{m}}\!=\!\tfrac{\omega}{c}\tfrac{\epsilon_{\mathrm{m}}(\omega)}{\sqrt{\epsilon_{\mathrm{d}}(\omega)+\epsilon_{\mathrm{m}}(\omega)}}$ in the metal, both of which take on imaginary values. We employ the Lorentz-Drude model \cite{Rakic98AO} for $\epsilon_{\mathrm{m}}(\omega)$ that combines the intraband contribution (the free-electron effects described by the Drude model) and the interband contribution (the bound-electron effects described by the Lorentz model for insulators). The SPP light-cone takes the form $k_{x}^{2}+k_{z}^{2}\!=\!k_{\mathrm{LD}}^{2}(\omega)$, where $k_{\mathrm{LD}}$ represents the real part of the SPP dispersion relation or light-line according to the dielectric functions of the materials described above. A pulsed SPP of finite transverse width corresponds to a 2D spectral support domain on the surface of the SPP light-cone [Fig.~\ref{Fig:LightConeRepresentation}(c)]. We define a ST-SPP \cite{Schepler20ACSP} as that surface-bound wave packet whose spectral support domain is the 1D trajectory at the intersection of the SPP light-cone with a spectral plane $k_{z}\!=\!k_{\mathrm{o}}'+\Omega/\widetilde{v}$ [Fig.~\ref{Fig:LightConeRepresentation}(d)]; where $\widetilde{v}$ is its group velocity, and $k_{\mathrm{o}}'\!\neq\!k_{\mathrm{o}}$ is the SPP wave number on the SPP light-line ($k_{x}\!=\!0$ in the SPP light-cone) evaluated at $\omega_{\mathrm{o}}$. Such a ST-SPP travels rigidly along the metal-dielectric interface without diffraction or dispersion -- independently of the material parameters -- at a group velocity $\widetilde{v}$ that can in principle take on arbitrary values \cite{Schepler20ACSP}.

\begin{figure*}[t!]
  \begin{center}
  \includegraphics[width=16.5cm]{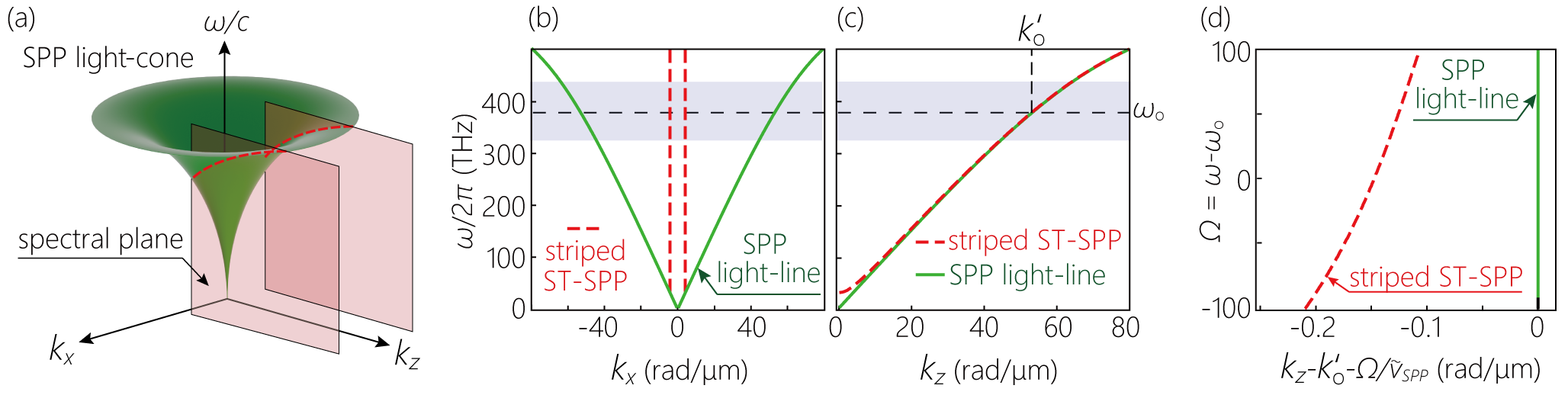}
  \end{center}
  \caption{The spectral representation of striped ST-SPPs. (a) Schematic of the SPP light-cone in $(k_{x},k_{z},\tfrac{\omega}{c})$ spectral space. The spectral support domain for a striped ST-SPP is the intersection of the SPP light-cone with two vertical iso-$k_{x}$ planes corresponding to $\lambda_{x}\!=\!\tfrac{2\pi}{k_{x}}$. (b) Projection of the spectral support domain for the striped ST-SPP onto the $(k_{x},\omega)$-plane and (c) onto the $(k_{z},\omega)$-plane. The translucent blue band in (b) and (c) corresponds to the bandwidth used in our experiments, centered at $\tfrac{\omega_{\mathrm{o}}}{2\pi}\!\approx\!375$~THz. (d) Because the SPP light-line and the striped ST-SPP dispersion curves in (c) almost coincide, we re-plot (c) here after modifying the horizontal axis to $k_{z}-k_{\mathrm{o}}'-\Omega/\widetilde{v}_{\mathrm{SPP}}$ to highlight the difference between the two curves.}
  \label{Fig:StripedTheory}
\end{figure*}

In our initial experiments reported here, we have made use of ultrabroadband pulses of bandwidth $\Delta\lambda\!=\!110$~nm (FWHM) at a center wavelength of $\lambda_{\mathrm{o}}\!=\!800$~nm. Because the spatial and temporal bandwidths of a ST-SPP are related to each other, just as in the case of free-space ST wave packets, the large $\Delta\lambda$ employed necessitates an extremely large $\Delta k_{x}$, thereby requiring operation deep within the non-paraxial regime. To avoid such an exorbitant requirement at this early stage of development of ST-SPPs, we instead consider a spectral support domain in which $|k_{x}|\!=\!\tfrac{2\pi}{\lambda_{x}}$ is held constant [Fig.~\ref{Fig:StripedTheory}(a)], where $\lambda_{x}$ is a transverse length scale. Therefore, $\sin\{\varphi(\omega)\}\!=\!\tfrac{2\pi c}{\lambda_{x}\omega}$, so that $\varphi(\lambda)\!\approx\!\tfrac{\lambda}{\lambda_{x}}$ in the small-angle approximation. By maintaining the linear proportionality between $\varphi$ and $\lambda$, we can exploit the full bandwidth of $\Delta\lambda\!\sim\!110$~nm within the paraxial domain ($\varphi\!<\!5^{\circ}$). The fixed $k_{x}$ entails a periodic transverse spatial profile for the ST-SPP of period $\lambda_{x}$. Note that $\lambda_{x}$ is \textit{not} the laser wavelength, but is instead a transverse spatial period characterizing the field structure. We thus call this structured surface wave a striped ST-SPP.

The spectral support domain of this striped ST-SPP [Fig.~\ref{Fig:StripedTheory}(a)] is the intersection of the SPP light-cone with the iso-$|k_{x}|$ planes. The spectral projection onto the $(k_{x},\tfrac{\omega}{c})$-plane takes the form of two vertical lines $k_{x}\!=\!\pm\tfrac{2\pi}{\lambda_{x}}$ [Fig.~\ref{Fig:StripedTheory}(b)], and that onto the $(k_{z},\tfrac{\omega}{c})$-plane takes the form of a curve that is close to the SPP light-line [Fig.~\ref{Fig:StripedTheory}(c)]. To delineate the SPP light-line from the striped-ST-SPP dispersion curve, we re-plot in Fig.~\ref{Fig:StripedTheory}(d) the spectral projection from Fig.~\ref{Fig:StripedTheory}(c) after transforming the horizontal axis $k_{z}\!\rightarrow\!k_{z}-k_{\mathrm{o}}'-\tfrac{\Omega}{\widetilde{v}_{\mathrm{SPP}}}$, where $\widetilde{v}_{\mathrm{SPP}}$ is the group velocity of the SPP evaluated at $\omega_{\mathrm{o}}$ (the slope of the SPP light-line at $\omega_{\mathrm{o}}$). The SPP light-line in this case becomes a vertical line at $k_{z}-k_{\mathrm{o}}'-\tfrac{\Omega}{\widetilde{v}_{\mathrm{SPP}}}\!=\!0$, and the spectral projection for the striped ST-SPP is well-separated from it [Fig.~\ref{Fig:StripedTheory}(d)]. This is the spatio-temporal spectral structure that must be inculcated into the SPP to yield a diffraction-free striped ST-SPP.

\subsection{Simulations}

\begin{figure*}[t!]
  \begin{center}
  \includegraphics[width=16.5cm]{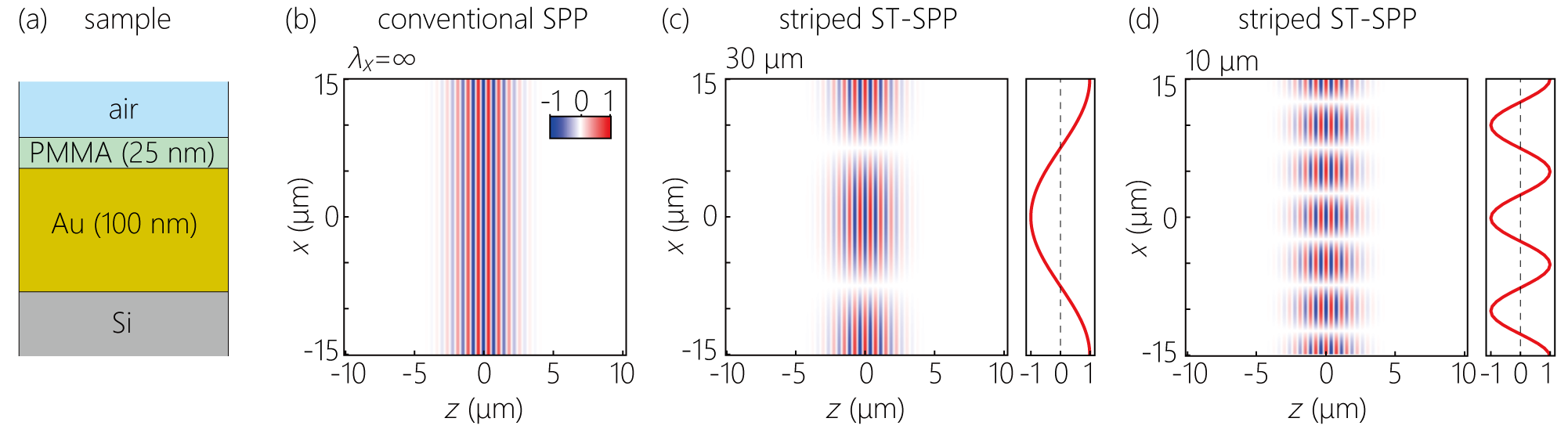}
  \end{center}
  \caption{Calculated out-of-plane electric field distributions $E_{y}(x,z;t=0)$ for striped ST-SPPs. (a) Structure of the sample: a 100-nm Au film deposited on a Si substrate, followed by a 25-nm PMMA film, and then air. (b) Calculated field distribution $E_{\mathrm{SPP}}(x,z;t\!=\!0)$ for a pulsed plane-wave SPP excited by a transform-limited laser pulse, corresponding to a striped ST-SPP with infinite $\lambda_{x}$. (c,d) Same as (b) for striped ST-SPPs, $E_{\mathrm{striped}}(x,z;t\!=\!0)$, with (c) $\lambda_{x}\!=\!30$~$\mu$m and (d) $\lambda_{x}\!=\!10$~$\mu$m. The striped ST-SPP in (d) corresponds to the spectral support domain in Fig.~\ref{Fig:StripedTheory}. The red curves in the side panels in (c) and (d) are cross sections through the field distributions in the main panels at $z\!=\!0$.}
  \label{Fig:Simulations}
\end{figure*}

We plot in Fig.~\ref{Fig:Simulations} calculated profiles of striped ST-SPPs for the sample of interest, which comprises a 100-nm-thick Au film deposited on a Si substrate, provided with a 25-nm-thick layer of poly(methyl methacrylate) (PMMA), followed by air [Fig.~\ref{Fig:Simulations}(a)]. The dispersion curve of the sample surface was calculated using a model proposed by Pockrand \cite{Pockrand78SS}, employing the dielectric function for Au proposed by Raki{\'c} \textit{et al}. \cite{Rakic98AO}.

We first calculate the out-of-plane field $E_{y}(x,z;t)$ for a conventional plane-wave pulsed SPP. The free-space pulse used to excite the SPP has the form $E(t)\!=\!E_{\mathrm{o}}e^{-i\omega_{\mathrm{o}}t}\exp{(-\tfrac{t^{2}}{(\Delta T/2\ln{2})^{2}})}$, where $\Delta T\!=10$~fs and $\tfrac{\omega_{\mathrm{o}}}{2\pi}\!=\!375$~THz. Obtaining the pulse spectrum $\widetilde{E}(\omega)$, we calculate the SPP field $E_{\mathrm{SPP}}(x,z;t)\!\propto\!\int\!d\omega\widetilde{E}(\omega)e^{i(k_{z}(\omega)z-\omega t)}$, where $k_{z}(\omega)$ and $\omega$ are related through the SPP dispersion relation. We plot $E_{\mathrm{SPP}}(x,z;t\!=\!0)$ in Fig.~\ref{Fig:Simulations}(b); the transverse extent along $x$ is infinite, and along $z$ is $\approx\!2.1$~$\mu$m. The field for the striped ST-SPPs is given by $E_{\mathrm{striped}}(x,z;t)\!\propto\!\cos{k_{x}x}\int\!d\omega\widetilde{E}(\omega)e^{i(k_{z}(k_{x},\omega)z-\omega t)}$, where $k_{z}(k_{x},\omega)\!=\!\sqrt{k_{\mathrm{LD}}^{2}(\omega)-k_{x}^{2}}$. We plot $E_{\mathrm{striped}}(x,z;t\!=\!0)$ in Fig.~\ref{Fig:Simulations}(c) for $\lambda_{x}\!=\!30$~$\mu$m, and in Fig.~\ref{Fig:Simulations}(d) for $\lambda_{x}\!=\!10$~$\mu$m. The striped ST-SPP is localized in the out-of-plane dimension $y$ by virtue of plasmonic confinement, is confined along $z$ [Fig.~\ref{Fig:Simulations}(b-d)] because of the finite pulse duration, and is periodic along the transverse coordinate $x$. Combining striped ST-SPPs of different spatial frequencies $k_{x}$ produces a ST-SPP that is also localized along $x$ \cite{Schepler20ACSP}.

\section{Experimental arrangement and measurement results}

\subsection{Free-space synthesis of ST wave packets}

\begin{figure}[t!]
  \begin{center}
  \includegraphics[width=8.6cm]{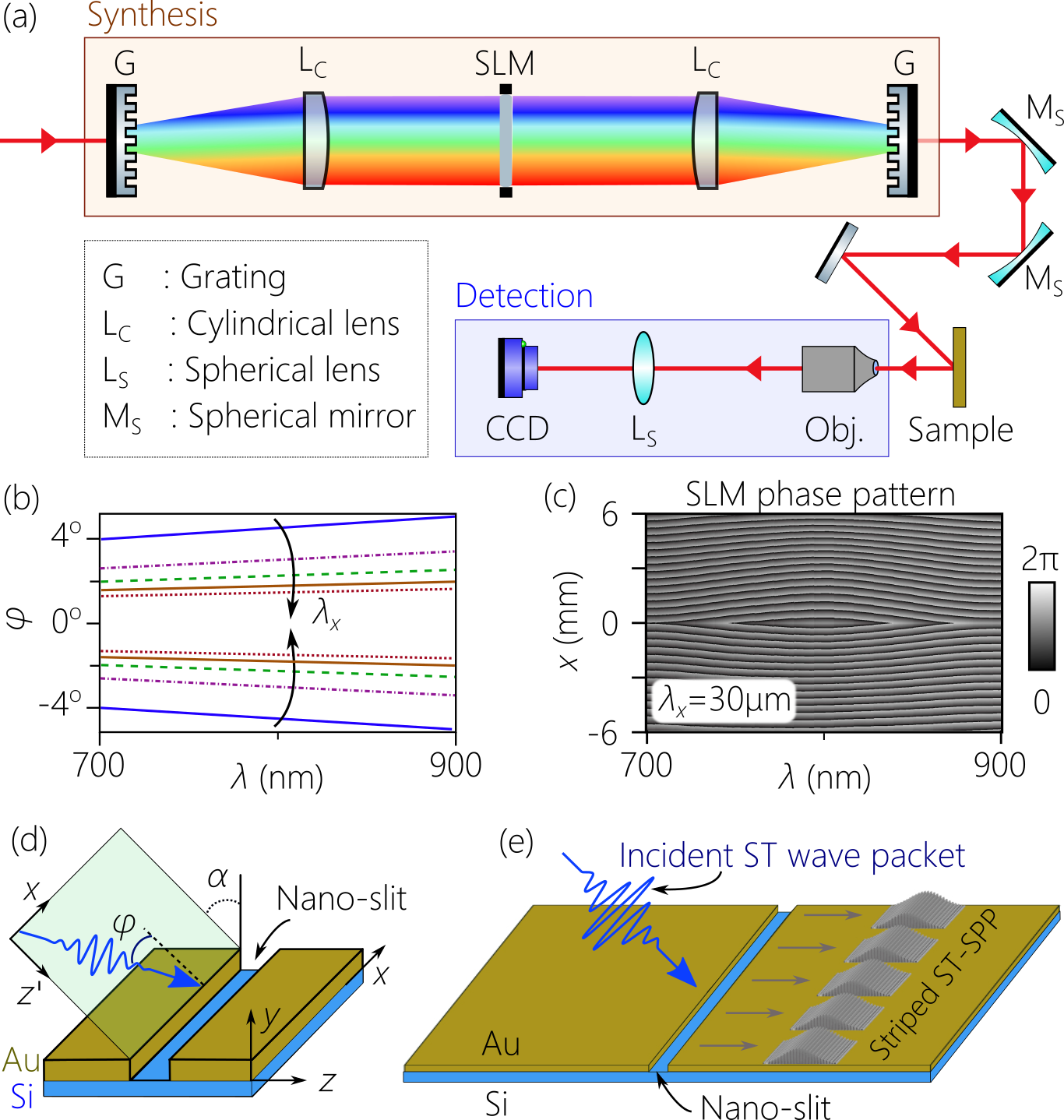}
  \end{center}
  \caption{Experimental arrangement for launching striped ST-SPPs from free-space ST wave packets. (a) Schematic of the entire arrangement, including synthesis of free-space ST wave packets, excitation of striped ST-SPPs, and detection. The SLM used in the experiment was in reflection mode, but we depict an unfolded configuration for clarity. (b) The deflection angle $\pm\varphi(\lambda)$ imparted by the SLM to each incident wavelength $\lambda$ to yield a striped ST-SPP of spatial period $\lambda_{x}$ extending from 10~$\mu$m to 30~$\mu$m. (c) SLM phase pattern used to produce the ST wave packet in free space (for $\lambda_{x}\!=\!30$~$\mu$m). (d,e) Schematic of the sample geometry (gold film on a Si substrate) including the nano-slit structure. The Au surface is covered with a 25-nm-thick dye-doped PMMA film. The plane of the incident ST wave packet is inclined by an angle $\alpha\!=\!50^{\circ}$ with respect to the surface normal, and is parallel to the nano-slit.}
  \label{Fig:Setup}
\end{figure}

The overall experimental arrangement for synthesizing and characterizing striped ST-SPPs is shown in Fig.~\ref{Fig:Setup}(a). The first step comprises synthesizing ST wave packets in free space from a generic pulsed beam via a well-established approach \cite{Kondakci17NP,Kondakci19NC}. The light source is a custom-made Ti:sapphire laser oscillator with a transform-limited pulse duration of 10~fs, a spectrum ranging from 680~nm to 900~nm (center wavelength $\lambda_{\mathrm{o}}\!=\!800$~nm and FWHM-bandwidth $\Delta\lambda\!=\!110$~nm), a repetition rate of 90~MHz, and an average power of 450~mW. Because of residual chirp in the optical system, the pulses are not transform limited, and instead have a pulse duration of $\Delta T\!\approx\!16$~fs at the sample surface. The spectrum of these femtosecond pulses is spatially resolved by a grating (300~lines/mm), collimated by a cylindrical lens L$_{\mathrm{c}}$ (focal length $f\!=\!250$~mm), and directed to a 2D phase-only SLM (Hamamatsu X13138-07). Each wavelength occupies a column on the SLM active area. The SLM imparts a phase distribution in each column to deflect the wavelength $\lambda$ by an angle $\pm\varphi(\lambda)$, which corresponds to a fixed spatial frequency $k_{x}\!=\!\tfrac{2\pi}{\lambda}\sin\{\varphi(\lambda)\}\!=\!\tfrac{2\pi}{\lambda_{x}}$ [Fig.~\ref{Fig:Setup}(b,c)]. The initial wavefront is incident on the SLM at an angle $\approx\!3^{\circ}$, and the reflected phase-modulated wavefront is directed to a second grating (identical to the one in the path of the incident field), whereupon the pulse is reconstituted to produce the ST wave packet. 

\subsection{Coupling from a free-space ST wave packet to a ST-SPP}

The metal surface we make use of is a 100-nm-thick Au film thermally evaporated onto a Si wafer. A 25-nm-thick PMMA film is spin-coated on the Au surface. The PMMA film is doped with a laser dye (Coumarin 343) to form a fluorescent layer. To reduce the frequency dependence of the coupling efficiency, a single 100-nm-wide slit \cite{Kubo07NL,Zhang11PRB,Zhang13JPCC} is milled into the Au surface via a focused ion beam [Fig.~\ref{Fig:Setup}(d)]. The $(x,z')$-plane for the incident free-space ST wave packet is parallel to the nano-slit and makes an angle $\alpha\!=\!50^{\circ}$ with respect to the $(x,z)$-plane normal to the Au surface covering an area $\approx\!150\times200$~$\mu$m$^2$ [Fig.~\ref{Fig:Setup}(d)]. Because the nano-slit has translational symmetry along the $x$-direction, the continuity of the wavefront along $x$ is conserved in the process of coupling the incident free-space field to a SPP. Therefore, the free-space ST wave packet is coupled to a striped ST-SPP having the same $k_{x}$ [Fig.~\ref{Fig:Setup}(e)]. 

\begin{figure*}[t!]
  \begin{center}
  \includegraphics[width=14.6cm]{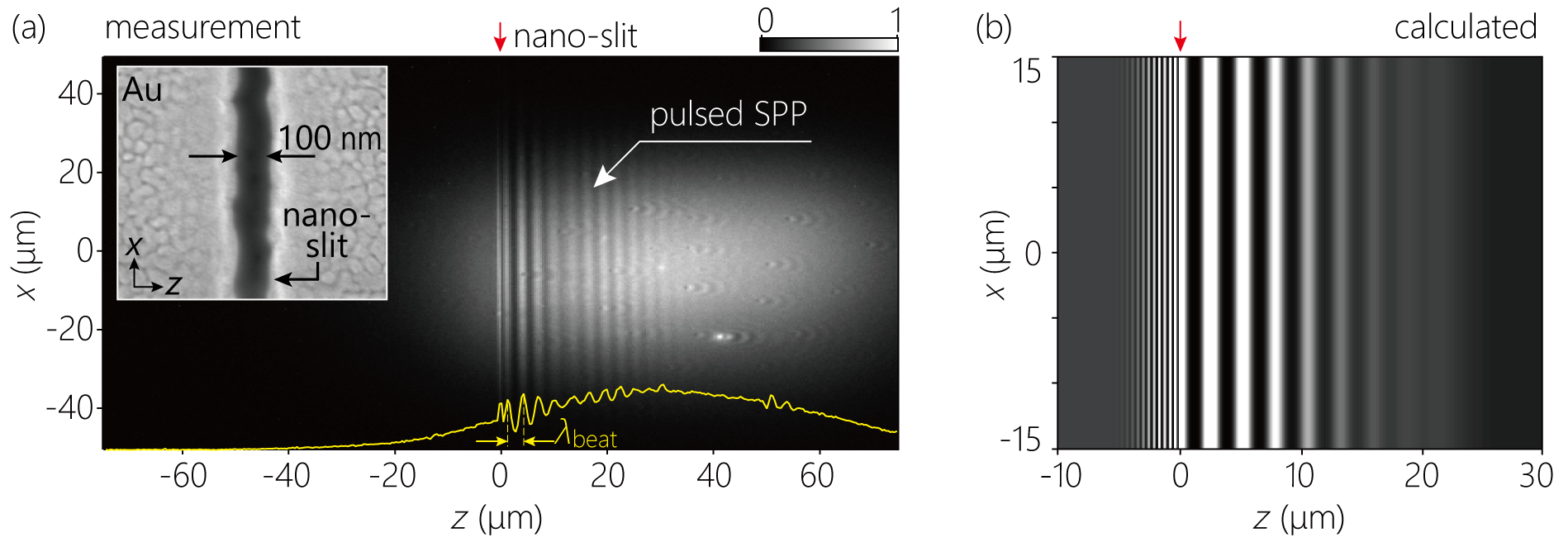}
  \end{center}
  \caption{(a) Optical micrograph in the $(x,z)$-plane of a conventional pulsed SPP excited via the nano-slit. An interference pattern with axial beating length $\lambda_{\mathrm{beat}}$ is observed because of two-photon fluorescence produced by the superposition of the free-space and SPP fields. The yellow curve at the bottom is the detected fluorescence signal along the $z$-axis after integrating along $x$. The red arrow at the top indicates the location of the nano-slit. The inset is a SEM micrograph of the nano-slit structure. (b) Calculated intensity profile $I(x,z)$ of the time-averaged two-photon fluorescence produced by the superposition of a plane-wave pulsed incident field and the excited SPP.}
  \label{Fig:BasicObservation}
\end{figure*}

\subsection{Detection of conventional SPPs}

The excited striped ST-SPP and the incident free-space ST wave packet co-exist at the Au surface covered with the dye-doped PMMA film. The mutual coherence between the free ST wave packet and the surface-bound SPP leads to an interference pattern with an axial beat length along the $z$-axis given by 
\begin{equation}\label{Eq:BeatPeriod}
\lambda_{\mathrm{beat}}=\frac{2\pi}{|k_{\mathrm{SPP}}-k_{\mathrm{o}}\sin\alpha|},
\end{equation}
where $k_{\mathrm{SPP}}\!=\!k_{z}(\omega_{\mathrm{o}},k_{x})\!\approx\!k_{\mathrm{o}}'$ is the SPP wave number, and $k_{\mathrm{o}}\sin{\alpha}$ is the in-plane wave number of the incident field. We detect the time-averaged two-photon fluorescence signal $I(x,z)$ that retains this beat length, where:
\begin{equation}\label{Eq:2PIntensity}
I(x,z)\propto\int\!dt\;\left(\left|E_{\mathrm{SPP}}(x,z;t)+E_{\mathrm{free}}(x,z;t)\right|^{2}\right)^{2};
\end{equation}
here, $E_{\mathrm{SPP}}$ and $E_{\mathrm{free}}$ are the scalar representations of the excited SPP and the free-space electric fields in the PMMA film, respectively. Typically, the out-of-plane component dominates the field intensity. The two-photon-fluorescence emission is collected with an objective lens (M Plan Apo SL20X, Mitutoyo) equipped with a band-pass filter transmitting light in the range $475-495$~nm (Semrock, FT-02-485/20-25), followed by a CCD camera (Rolera EM-C2, QImaging) \cite{Hattori12JJAP,Ichiji22NanoP}. The intensity $I(x,z)$ detected by the CCD camera is a time integral of the fourth power of the total electric field.

Figure~\ref{Fig:BasicObservation}(a) shows an optical micrograph of the two-photon-fluorescence signal for a conventional pulsed SPP (after setting the SLM phase to 0). As shown in Fig.~\ref{Fig:Setup}(d,e), the plane of the incident field is tilted with respect to the normal to the sample. The incident field launches SPPs from the nano-slit located at $z\!\approx\!0$ in the forward (to the right in Fig.~\ref{Fig:BasicObservation}(a), $z\!>\!0$) and backward (to the left in Fig.~\ref{Fig:BasicObservation}(a), $z\!<\!0$) directions. The backward-coupled SPP is weaker, and the associated axial beating length is $\lambda_{\mathrm{beat}}\!\approx\!0.5$~$\mu$m (according to Eq.~\ref{Eq:BeatPeriod}), which is finer than the spatial resolution of our imaging system. The forward-coupled SPP has $\lambda_{\mathrm{beat}}\!\approx\!2.8$~$\mu$m. In our measurements of striped ST-SPPs, we focus exclusively on the forward-coupled SPP propagating to the right $z\!>\!0$ in Fig.~\ref{Fig:BasicObservation}(a).

The overall spatial distribution of the two-photon-fluorescence emission is impacted by the shape of the excitation-laser spot on the sample surface. In Fig.~\ref{Fig:BasicObservation}(a), the fluorescence intensity has its maximum at $z\!\approx\!30$~$\mu$m because the center of the laser spot is located there. Nevertheless, in the vicinity of $z\!=\!0$, the visibility of the interference pattern is large, indicating comparable contributions from $E_{\mathrm{SPP}}$ and $E_{\mathrm{free}}$ in Eq.~\ref{Eq:2PIntensity}. This is because the signal is obtained from the 25-nm-thick PMMA film, and the \textit{intensity} of the surface bound SPP is very high due to field localization at the Au surface.

We plot in Fig.~\ref{Fig:BasicObservation}(b) a calculated spatial profile in the $(x,z)$-plane of the expected two-photon-fluorescence intensity distribution resulting from the interference of an incident free-space field and the excited SPP. We make use of Eq.~\ref{Eq:2PIntensity} and assume a plane-wave pulsed incident field. The calculation is repeated for the forward- and backward-coupled SPPs, which are characterized by different axial beat lengths $\lambda_{\mathrm{beat}}$. The calculated intensity profile [Fig.~\ref{Fig:BasicObservation}(b)] is in excellent agreement with the measured distribution [Fig.~\ref{Fig:BasicObservation}(a)].

\subsection{Detection of striped ST-SPPs}

\begin{figure}[t!]
  \begin{center}
  \includegraphics[width=8.6cm]{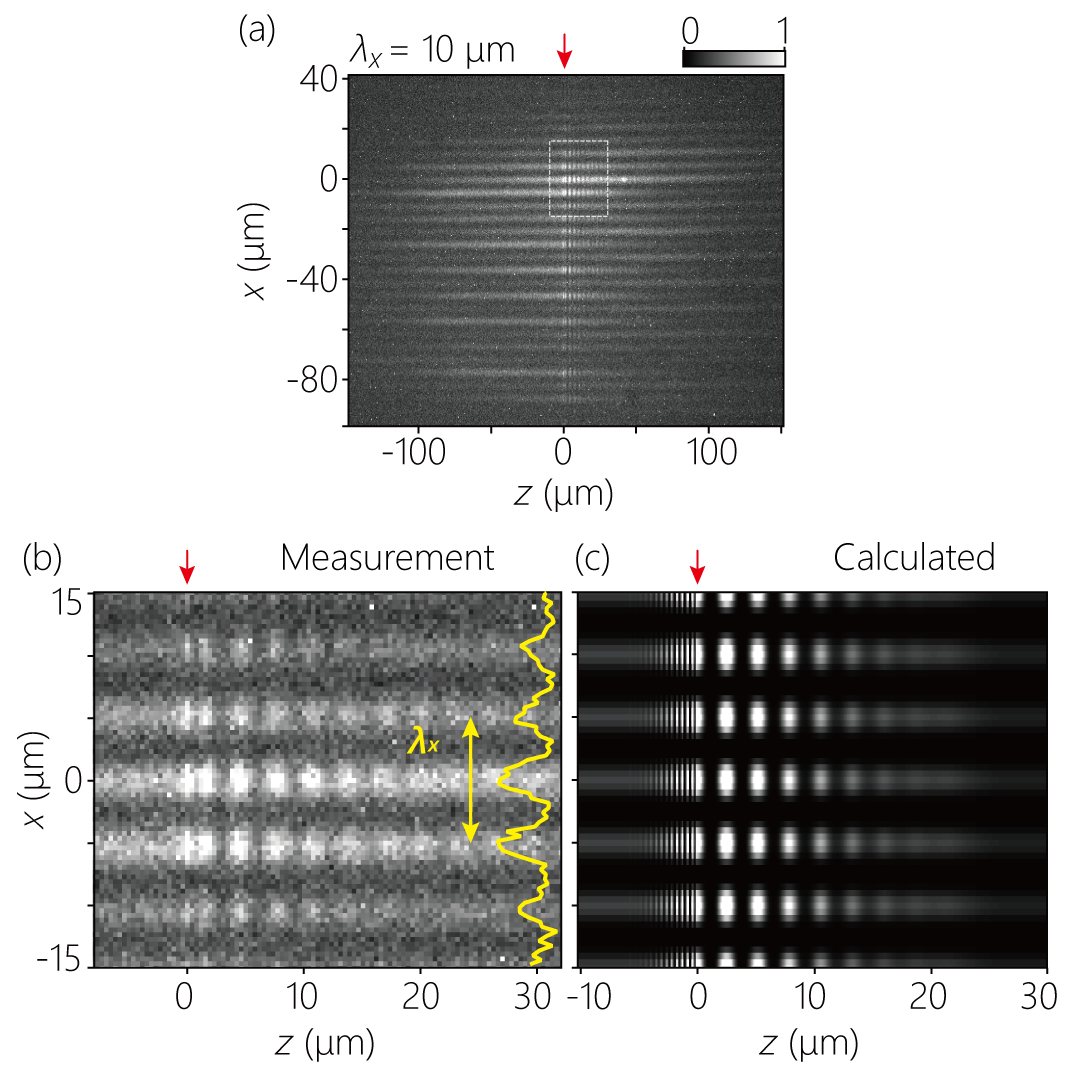}
  \end{center}
  \caption{Measured and calculated intensity of the two-photon fluorescence produced by striped ST-SPPs. (a) Micrograph of two-photon fluorescence microscopy for a striped ST-SPP with $\lambda_{x}\!=\!10$~$\mu$m. The dotted box is the area expanded in (b). (b) Expanded view of (a) corresponding to the dotted box. The yellow curve on the right results from integration along $z$. (c) Calculated field distribution of the two-photon excitation, which results from the interference between the striped ST-SPP and the incident free-space wave packet when $\lambda_{x}\!=\!10$~$\mu$m. The red arrows at the top of the panels identify the location of the nano-slit at $z\!\approx\!0$.}
  \label{Fig:DataTheory}
\end{figure}

We now excite the sample with the ST wave packet having constant $|k_{x}|$. Figure~\ref{Fig:DataTheory}(a) is an optical micrograph of the two-photon fluorescence obtained with an incident ST wave packet with $\lambda_{x}\!=\!10$~$\mu$m. A magnified view in Fig.~\ref{Fig:DataTheory}(b) shows the axial beat pattern excited after the nano-slit (located at $z\!\approx\!0$) couples the incident field to a forward-coupled surface-bound wave packet propagating to the right $z\!>\!0$). Here, in addition to the axial beating observed for the pulsed SPP in Fig.~\ref{Fig:BasicObservation}(a), we observe an additional field structure: a transverse periodic spatial structure with period $\lambda_{x}/2\!=\!5$~$\mu$m. The halving of the transverse period in the micrograph compared to $\lambda_{x}$ simply reflects the structure of the ST-SPP field containing two phase-inverted antinodes within one transverse cycle. The measured axial beating length is $\lambda_{\mathrm{beat}}\!\approx\!2.9$~$\mu$m. Figure~\ref{Fig:DataTheory}(c) shows the calculated intensity profile for the two-photon fluorescence signal resulting from the interference of the striped ST-SPP with the incident ST wave packet. The calculation is performed for both the forward- and backward-coupled striped ST-SPPs using the same approach employed in Fig.~\ref{Fig:BasicObservation}(b).

\begin{figure}[t!]
  \begin{center}
  \includegraphics[width=8.6cm]{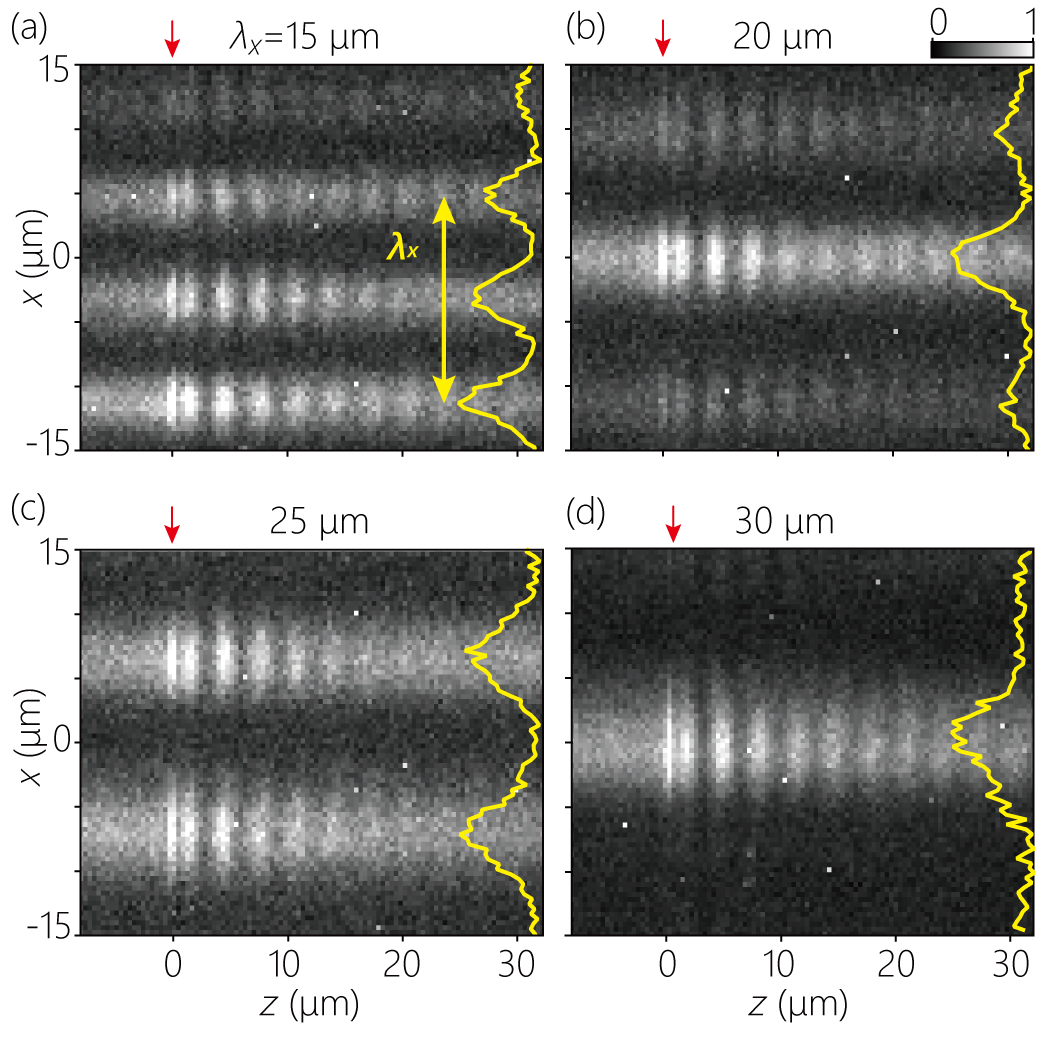}
  \end{center}
  \caption{Optical micrograph of the two-photon fluorescence for striped ST-SPPs with (a) $\lambda_{x}\!\approx\!15$~$\mu$m, (b) $20$~$\mu$m, (c) $25$~$\mu$m, and (d) $30$~$\mu$m. The red arrow on the top of each panel identifies the location of the nano-slit at $z\!\approx\!0$. The axial beat length $\lambda_{x}$ is approximately constant. The yellow curves on the right of each panel is the transverse intensity profile resulting from integration along $z$.}
  \label{Fig:ChangingWidth}
\end{figure}

By simply modifying the 2D phase distribution imparted by the SLM to the spectrally resolved wave front of the free-space pulse [Fig.~\ref{Fig:Setup}(c)], we can tune the proportionality between $\varphi$ and $\lambda$ [Fig.~\ref{Fig:Setup}(b)], thus producing a different value for $k_{x}$. Once this new ST wave packet is coupled to a SPP via the nano-slit, a striped ST-SPP is formed having a new transverse period $\lambda_{x}$. We plot in Fig.~\ref{Fig:ChangingWidth} measurements of four such striped ST-SPPs with $\lambda_{x}$ increased in 5-$\mu$m steps from $\lambda_{x}\!=\!15$~$\mu$m [Fig.~\ref{Fig:ChangingWidth}(a)] to $\lambda_{x}\!=\!30$~$\mu$m [Fig.~\ref{Fig:ChangingWidth}(d)]. The axial beat length changes only slightly with the tuning of $k_{x}$: it is largest for the smallest $k_{x}$ (or largest $\lambda_{x}$).

An important observation can be made based on Fig.~\ref{Fig:DataTheory}(b) and Fig.~\ref{Fig:ChangingWidth}(a-d). The measured transverse period $\lambda_{x}$ in each of these (from $\lambda_{x}\!=\!10$~$\mu$m to 30~$\mu$m) matches that expected from the exited value of $k_{x}$, and no extra transverse beating is observed over the observation area on the Au surface. The absence of addition transverse beating indicates that the transverse period of the ST-SPP matches that of the incident ST wave packet. Therefore, by combining a multiplicity of spatial frequencies $k_{x}$ in the incident ST wave packet to produce a localized peak along $x$, a corresponding ST-SPP similarly localized along $x$ -- and that is, furthermore, diffraction-free -- is to be expected.

\section{Discussion and Conclusion}

The results reported here emphasize the possibilities emerging from interfacing the rapidly developing topic of ST wave packets \cite{Yessenov22AOP} with nanophotonics. Recent efforts along these lines include exploiting nanophotonic structures for synthesizing ST wave packets \cite{Guo21Light}, in addition to the propagation of ST wave packets in waveguides \cite{Shiri20NC_Hybrid,Kibler21PRL,Guo21PRR,Ruano21JO,Bejot2021ACSP,Shiri22ACSP}.

Although we have produced here a particular example of a ST-SPP (namely, a striped ST-SPP), the experimental arrangement we have constructed is capable of synthesizing a variety of other ST-SPPs (such as those in Ref.~\cite{Schepler20ACSP}) by changing the SLM phase pattern employed [Fig.~\ref{Fig:Setup}(a,c)], which will be the focus of our future efforts. The bandwidth exploited here $\Delta\lambda\!=\!110$~nm is the largest for a ST wave packet to date, exceeding the previous record \cite{Kondakci18OE} of $\Delta\lambda\!=\!30$~nm. Moreover, to retain this large bandwidth while localizing the ST-SPP along $x$ in the paraxial regime, we can exploit the recently demonstrated strategy of `spectral recycling' \cite{Hall21PRArecycling}. Finally, our current detection scheme unveils the time-averaged intensity profile of the ST-SPP. Reconstructing the unique X-shaped spatio-temporal profiles of ST-SPPs [Fig.~\ref{Fig:GeneralConcept}(a,b)] requires time-resolved measurements, which we are currently constructing. Furthermore, such a time-resolved detection scheme is required to measure the group velocity of the ST-SPP on the Au surface, which is predicted to be tunable over a wide span of values \cite{Schepler20ACSP}. 

An intriguing possibility emerges from our recent experimental demonstration of accelerating ST wave packets in free space with record-high magnitudes of acceleration \cite{Yessenov20PRL2} ($\sim\!4-5$ orders-of-magnitudes larger than previous reports \cite{Clerici08OE,Lukner09OE,Chong10NP}), in addition to arbitrary acceleration profiles along the propagation axis \cite{Hall22OLaccel}. These advances suggest the intriguing potential for exciting \textit{accelerating} ST-SPPs, which may lead to new forms of radiation produced from the accelerated surface plasmons \cite{Henstridge18Science}.

Because the axial wave number $k_{z}$ of the excited striped ST-SPP is \textit{not} linear in $\Omega$ (because $k_{x}$ is constant), striped ST-SPPs are dispersive. Indeed, they correspond to the wave packets in \cite{Liu98JMO,Hu02JOSAA,Lu03JOSAA,Giovannini15Science} known as pulsed Bessel beams in which a fixed radial wave number $k_{r}$ is maintained constant, resulting in a diffraction-free Bessel beam structure as the transverse spatial profile, but a temporally dispersive pulse structure. Because SPPs are surface waves that have only one transverse dimension, the Bessel profile degenerates into a cosine profile.

In conclusion, we have provided the first experimental evidence for ST-SPPs and the possibility of efficiently exciting them at a gold-dielectric interface. Starting with a 16-fs pulse at a wavelength of 800~nm, we form a ST wave packet in free space that retains the full pulse bandwidth (FWHM $\Delta\lambda\!=\!110$~nm) and couple it to a SPP via a 100-nm-wide nano-slit milled into the Au surface. As a first step, we have made use of a ST wave packet in which a single spatial frequency $k_{x}\!=\!\tfrac{2\pi}{\lambda_{x}}$ is exploited (the propagation angle of each frequency with respect to a common direction is linear in frequency). This free-space wave packet is coupled by a nano-slit to a propagation-invariant ST-SPP whose transverse spatial profile is periodically modulated (or striped) with a period $\lambda_{x}$. We have produced and observed striped ST-SPPs with $\lambda_{x}$ in the range from 10~$\mu$m to 30~$\mu$m by changing the 2D phase pattern imparted by a SLM to the spatially resolved spectrum of the initial generic laser pulse in free space. Future work will incorporate a continuous spatial spectrum rather than a single spatial frequency, which yields a localized rather than periodic spatial profile; time-resolved measurements to verify the propagation invariance of the spatio-temporal profile for the ST-SPP; and tuning of the group velocity of the excited surface plasmon.

\section*{Acknowledgments}
The authors thank K.~Oshima for contributions to the fabrication of the samples. This work was supported by the JSPS KAKENHI (JP16823280, JP18967972, JP20J21825); MEXT Q-LEAP ATTO (JPMXS0118068681); Nanotechnology Platform Project (JPMXP09F-17-NM-0068); by the Nanofabrication Platform of NIMS and University of Tsukuba; and by the U.S. Office of Naval Research (ONR) under contracts N00014-17-1-2458 and N00014-20-1-2789.

\bibliography{diffraction}

\end{document}